\documentclass[preprint,authoryear,12pt]{elsarticle}

\usepackage{graphics}

\usepackage{amssymb}

\newcommand{\apj}{ApJ}
\newcommand{\apjl}{ApJ}
\newcommand{\apjs}{ApJS}
\newcommand{\nat}{Nature}
\newcommand{\aap}{A\&A}
\newcommand{\mnras}{MNRAS}
\newcommand{\aj}{AJ}
\newcommand{\araa}{ARA\&A}
\newcommand{\apss}{Ap\&SS}

\journal{Planetary and Space Science}

\begin{document}

\begin{frontmatter}


\title{Searching for Debris Disks around Isolated Pulsars}


\author{Zhongxiang Wang}
\ead{wangzx@shao.ac.cn}
\address{Shanghai Astronomical Observatory, Chinese Academy of Sciences,
Shanghai 200030, China}

\begin{abstract}
Different pieces of observational evidence suggest the existence of disks
around isolated neutron stars. Such disks could be formed from supernova
fallback when neutron stars are born in core-collapse supernova explosions. 
Efforts have been made to search for disks around different classes of
pulsars, which include millisecond pulsars, young neutron star classes
(magnetars, central compact objects, and X-ray dim isolated neutron stars),
and regular radio pulsars. We review the main results from observations 
at wavelengths of from optical to sub-millimeter/millimeter. 
\end{abstract}


\end{frontmatter}


\section{Introduction}
\label{sec:intr}

It has long been suggested that isolated pulsars (PSRs) might have disks and
the existence of disks would affect evolution of pulsars and help explain 
some observed features in pulsars \citep{md81}. The discovery of a planetary
system around PSR B1257+12 \citep{wf92}, which actually was the
first discovered extrasolar planetary system, motivated numerous studies about
how such a planetary system could be formed. The proposed formation mechanisms
more or less involved 
the existence of an accretion/debris disk around the neutron star
(see \citealt{mh01} and references therein). To date, two additional 
pulsars, B1620$-$26 and J1719$-$1438, have been found with planetary systems, 
although the first is a triple system in the globular cluster M4 likely 
having been formed from a dynamical exchange interaction 
(\citealt{sig+03} and references therein),
and in the second the planet-mass companion is probably the leftover of a
carbon white dwarf having lost most its mass during the phase of
low-mass X-ray binary evolution \citep{bai+11}.
In addition to these three pulsars, \citet{sha+13} recently have shown that
the observed, long-term timing variations for PSR B1937+21 are possibly 
explained by considering an asteroid belt around the pulsar.

The above neutron stars are old (ages $> 10^8$ yrs), so-called recycled 
pulsars with fast, millisecond spin periods. 
For regular pulsars formed from core-collapse
supernovae for 10$^3$--10$^6$ yrs, rotational spin noise 
(i.e., timing noise) in them is several
orders of magnitude larger than that in millisecond pulsars (MSPs) 
(e.g., \citealt{sc10}). It is thus difficult to detect planetary bodies 
around them from pulsar timing.  However certain phenomena have been seen 
hinting the existence of accretion/debris disks. \citet{cs08} summarize 
four types of pulsar pulse emission variations: nulling, transient pulse 
emitting from so-called rotating radio 
transients (RRATs; \citealt{mcl+06}), subpulse drifting, and emission mode 
changing, and consider that these phenomena are caused by migration of 
circumpulsar debris material into the magnetospheres of pulsars. 
In individual pulsars, for example, the recent measurement of the second 
period derivative of the spin of PSR J1734$-$3333 suggests that 
the magnetic field of this pulsar is increasing \citep{esp+11}, or 
alternatively its spin properties might be affected by having an accretion 
disk \citep{cal+13}. It has also been suggested that because jets are generally
associated with accretion disks, young pulsars seen with jets might harbor 
accretion disks \citep{bp04}.

While it is not very clear how disks around MSPs would be formed
(e.g., \citealt{sha+13}), particularly for the planetary system case around 
B1257+12 \citep{mh01}, young pulsars are believed to possibly have disks due to 
supernova fallback \citep{che89}. During a core-collapse supernova explosion,
part of ejected material may fallback and if the material has sufficient
angular momentum, a disk might be formed around the newly born neutron star
\citep*{lwb91}. The existence of fallback disks has been suggested to be
the cause of the diversity of young neutron stars (\citealt{alp01}; 
\citealt*{ace13} and references therein). Unlike what was once thought---young 
radio pulsars like those in the Crab and Vela supernova remnants were
prototypical of newborn neutron stars, it has been realized that there are
classes of magnetars \citep{wt06}, central compact objects 
(CCOs) in young supernova remnants \citep*{pst04,del08}, and X-ray dim 
isolated neutron stars (XDINSs; \citealt{tur09, mer11}). 
The latter three classes of young
neutron stars are rather `quiet' at multiple wavelength regions from 
optical to radio: they generally do not have strong non-thermal emission 
and are not surrounded by any bright, pulsar-wind powered nebulae.

Neutron stars generally have non-thermal emission radiated from their 
magnetospheres, and sometimes a thermal component arising from their 
hot surfaces may be seen (e.g., \citealt{bec09}; note that the thermal 
emission can be dominant 
in some cases). The non-thermal emission can be described by a power law
with flux decreasing from X-ray to optical/IR wavelengths, while because 
of their $\sim$10$^6$~K surface temperature, the thermal component's 
Rayleigh-Jeans tail may be detectable at ultraviolet/optical wavelengths 
(e.g., \citealt{kap+11}).
As a result, among $\sim$2000 known neutron stars \citep{man+05}, 
only over 20 neutron stars have been detected at optical/IR wavelengths.
For a comparison, a disk would have thermal-like emission, and depending
on temperature, which is often assumed to be a function of disk radius,
the disk would be generally bright at optical/IR wavelengths \citep{phn00}.
Emission from putative disks would thus be distinguishable from that 
of neutron stars.
In addition, if a debris disk consists of cold, $\leq$100~K dust, 
it would also possibly be detectable at submillimeter (submm) and 
millimeter (mm) wavelengths \citep{pc94}.

Given all these reasons, searches for disks around neutron stars have been 
carried out with different telescopes. The goal of the searches
is to find thermal-like emission from neutron stars at wavelengths of from
optical to submm/mm, which is not expected to be radiated from neutron stars 
themselves.  In this paper, we review the current 
status of searches and provide a summary of the main results.
It should be note that since young neutron stars (for example
the magnetars that are covered in this paper) may exhibit
strong variability (e.g., \citealt{kas07}), and hence (nearly) simultaneous.
observations at multiple wavelengths are often required in order to identify
the source of emission.  

\section{Searches for Disks around Different Classes of Neutron Stars}
\label{search}
\subsection{Millisecond pulsars}
\label{subsec:msp}

Motivated by the discovery of planets around B1257+12, searches for debris
disks (particularly) around similar old pulsars were made 
(e.g., \citealt{vt93,ff96,koc+02,lf04}). The observations were carried
out with either first generation infrared space telescopes (namely 
the \textit{Infrared Astronomical Satellite} and the 
\textit{Infrared Space Observatory}) or ground-based telescopes, and 
sensitivities were very limited, generally around 100 mJy at mid-infrared
(MIR) wavelengths of $\sim$10--100 $\mu$m. With simplified assumptions 
for dust grains and pulsar-wind heating of dust, the estimated surrounding
dust mass would be lower than 30 $M_{\oplus}$, for example, for the case
of B1257+12 \citep{ff96}. Note that since approximately more than 60\% 
known MSPs are in a binary system \citep{man+05}, binary       
MSPs quite often were also included as the targets in the searches.

Launched in 2003, \textit{Spitzer Space Telescope} provided observing
capabilities of imaging and spectroscopy at MIR wavelengths with 
sensitivities of from $\mu$Jy to mJy. \citet{bry+06} reported their
search for debris material around B1257+12 with \textit{Spitzer} MIPS 
observations. The sensitivities were improved by 3 orders of magnitude
comparing to those of the previous searches. No IR emission was detected
and they concluded that an asteroid belt of 0.01 $M_{\oplus}$, similar 
to that in the solar system, cannot be ruled out.

Efforts were also made at submm/mm 
wavelengths \citep{pc94,gh00,lww04}, but comparing to those at IR wavelengths
with temperature $T\geq 300$~K dust as the targets, the searches aimed to
detect colder, $T\sim 30$~K dust material around nearby MSPs. The typical
sensitivities reached were $\sim$5~mJy at bands within 0.8--3.0~mm, and 
mass limits of $\leq 10$~$M_{\oplus}$ were obtained. Since the dust is 
optically thin at submm/mm, the mass limits are rather certain, not depending 
on disk structures which have to be assumed in optically thick
disk cases.
\begin{figure}[t]
\begin{center}
\includegraphics[scale=0.55]{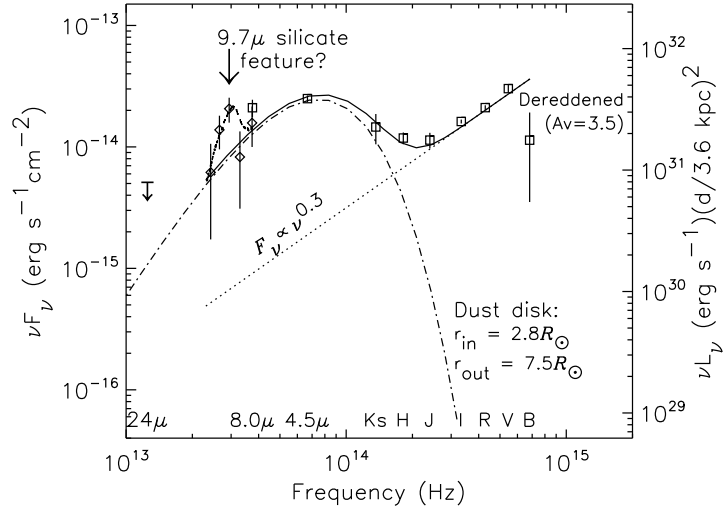}
\caption{Optical and IR broad-band and \textit{Spitzer} IRS spectrum 
of the magnetar 4U~0142+61 \citep*{wck08}. The squares are the optical, 
near-IR, \textit{Spitzer}\ IRAC 4.5/8.0 $\mu$m 
broadband fluxes (dereddened with $A_V=3.5$ mag), showing that the optical 
spectrum is consistent with being a power law (dotted line), 
$F_{\nu}\propto\nu^{0.3}$, and the IR spectrum can be fit with an X-ray 
irradiated dust disk model (dash-dotted curve). The diamonds are 
the dereddened IRS flux measurements, which appear as a bump when compared to
the dust disk model SED and can be fit with a silicate emission feature 
(dashed curve; \citealt{slo+03}).
The \textit{Spitzer} MIPS 24 $\mu$m upper limit is also included in the figure.}
\label{fig:sed}
\end{center}
\end{figure}

\subsection{Young neutron-star classes}
\subsubsection{Magnetars}
\label{mag}
Magnetars are considered to be young neutron stars with ultra-high, 
$\sim$10$^{14}$ to 10$^{15}$ Gauss surface magnetic fields, although
recent studies of the magnetars SGR 0418+5729 and J1822.3$-$1606, 
the surface magnetic fields of which were shown to be in the range of 
regular young radio pulsars (10$^{12}$--10$^{13}$ Gauss), challenge 
the conventional view \citep{rea+10,rea+12}. Around the year 2000,
detailed theoretical studies of fallback disks around young neutron stars,
particularly around magnetars and CCOs given their quietness at multiple
wavelength regions,
strongly suggested possible detections of them at optical and IR wavelengths 
\citep*{chn00,phn00,mph01}. A disk would be bright due to either internal 
viscous heating or irradiation by X-rays from the central neutron 
star. The discovery of the optical and near-IR (NIR) counterpart to
the magnetar 4U~0142+61 by \citet*{hvk00} seemed to have matched 
the theoretical expectations. However, followup optical timing of this magnetar 
by \citet{km02} found that its optical emission is pulsed at its spin
period with 27\% pulsed fraction. Such highly pulsed emission, when considering
4\%--14\% pulsed fraction in the magnetar's X-ray emission \citep{gon+10}, is
unlikely to originate from a disk.

In their systematic, deep search for disks around several magnetars and 
CCOs with the 
ground-based Magellan telescopes at optical and NIR wavelengths and 
\textit{Spitzer} at MIR wavelengths, the magnetar 4U~0142+61 was detected
by \citet*{wck06} at \textit{Spitzer} 4.5 and 8.0 $\mu$m bands. Combined
with the previous optical and NIR measurements, the spectral energy 
distribution (SED) of this magnetar from optical to MIR wavelengths was 
constructed and it likely contains two components: one a power-law spectrum 
over optical $V\/RI$ and NIR $J$-bands, probably arising from the 
magnetosphere of the pulsar, and one thermal blackbody-like over 
the 2.2--8~$\mu$m range, arising from a debris disk 
(\citealt*{wck06}; see also Figure~\ref{fig:sed}). 
The disk should probably have a size of 2.8~$R_{\odot}$ to 7.5~$R_{\odot}$,
estimated from fitting the IR component with an X-ray irradiated dust disk
model. Followup \textit{Spitzer} spectroscopy at 7.5--14~$\mu$m detected 
the magnetar with very low signal-to-noise ratios but imaging at 24~$\mu$m
did not, and the results from the both observations are consistent with
the dust disk model (\citealt*{wck08}; Figure~\ref{fig:sed}).

The magnetars are sources located at the Galactic plane, and thus extinctions
to them are large, making optical detections of them difficult. The magnetar
1E~2259+586 was the second one found with an NIR counterpart but only 
at $K_s$ band (2.1 $\mu$m; \citealt{hul+01}). After an X-ray outburst of
the source in 2002, its $K_s$ flux was observed to have an initial increase
and then decrease in concert with the X-ray flux \citep{tam+04}. On the basis
of the results, \citet{tam+04} suggested a neutron star magnetosphere origin 
for NIR emission, but it should be noted that the existence of a disk could
also be the link for such correlated flux variations, which was pointed out
and tested on 4U~0142+61 by \citet{wk08}. Deep \textit{Spitzer} MIR imaging
detected the counterpart to 1E~2259+586 at 4.5~$\mu$m band, and on the basis
of the two detections, overall IR emission from the magnetar was determined to
be similar to that from 4U~0142+61, suggesting the possible
existence of a debris disk \citep{kap+09}.

The magnetar 1E 1048.1$-$5937 was the third one discovered with an NIR 
counterpart \citep{wc02}, and the initial detection was likely 
associated with an X-ray flare of this magnetar in 2002 \citep{isr+02,dv05}.
No MIR counterparts were found with \textit{Spitzer} imaging at 4.5, 8.0, and
24 $\mu$m, although the derived flux upper limits were not very constraining
\citep*{wkh07}. In 2007, another X-ray flare was found occurring, with the 
X-ray flux approximately 3 times higher than that when the source was in 
the quiescent state \citep{tam+08}. Deep optical, NIR, and MIR observations 
during this flare
revealed that the optical and NIR fluxes were correlated with 
the X-ray flux, but
no MIR emission was detected \citep{wan+08}. The MIR flux upper limit at 
4.5~$\mu$m was sufficiently deep that similar MIR emission to that 
from 4U~0142+61 was excluded.

Magnetars are characterized by occasional high-energy outbursts or flares
(e.g., \citealt{kas07}),
and new members have thus been discovered. Optical and NIR observations
were often carried out once such outbursts/flares were identified by X-ray 
space telescopes. Among 21 confirmed magnetars (Mcgill SGR/AXP Online Catalog;
\citealt{ok13}), three other magnetars were confirmed with NIR counterparts
during their outbursts:
SGR~0501+4516 \citep{dhi+11}, SGR~1806$-$20 \citep{isr+05}, and 
XTE~J1810$-$197 \citep{cam+07,tes+08}. The NIR detections benefited from 
their emission brightening during outbursts,
and for SGR~0501+4516 a possible 20\% pulsed emission at $K$ band was 
detected \citep{dhi+11}.  The nature of the NIR emission is not clearly 
understood. In addition, the magnetars 1RXS~J170849.0$-$400910 and 
XTE~J1810$-$197 were also searched for their MIR counterparts with 
\textit{Spitzer}, but the observations were not sufficiently deep and
sub-mJy flux upper limits at 4.5~$\mu$m, 8.0~$\mu$m, and 24~$\mu$m were 
obtained (see Figure~2; \citealt*{wkh07}).
\begin{figure}[t]
\begin{center}
\includegraphics[scale=0.75]{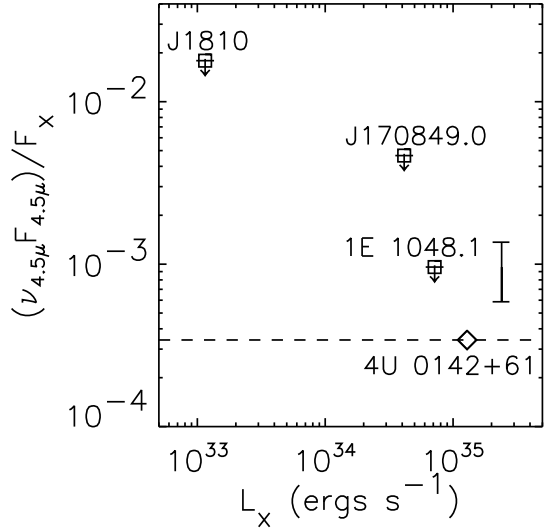}
\caption{Upper limits on the \textit{Spitzer}/IRAC 4.5~$\mu$m to X-ray 
(in the 2--10 keV range) flux ratios of the magnetars 1E~1048.1$-$5937,
1RXS~J170849.0$-$400910, andXTE~J1810$-$197. 
Fluxes of them are unabsorbed, and $A_V= 5.6, 7.7$, and 3.6 
are used for dereddening the 4.5~$\mu$m flux upper 
limits of the three magnetars, respectively. 
The error bar near 1E~1048.1$-$5937 indicates the uncertainty on its upper
limit due to its X-ray variability.
The upper limits are above the dashed line, which indicates the flux ratio 
of the magnetar 4U~0142+61, suggesting that the observations were not 
sufficiently deep.
}
\end{center}
\end{figure}

\subsubsection{CCOs}
\label{subsec:cco}
CCOs were probably the best neutron star candidates to be
searched for fallback disks.
They appear to have no non-thermal magnetospheric emission, no pulsar wind
nebulae, and be radio quiet, drastically different from Crab-pulsar--like young 
radio pulsars (see \citealt*{gha13} and references therein).
Deep optical and IR observations are therefore not interfered
by strong emission such as that from the Crab pulsar and its nebula, 
and are sensitive.
In addition, since they are located near the centers of 
young ($\sim$10$^{3}$ yr) supernova remnants (SNRs), their ages and distances 
are relatively certain. However deep searches have failed to detect 
optical/IR counterparts in any
of them \citep{fps06, wkc07, mig+07,mig+08,mig+09,del+11}. The obtained
optical and NIR upper limits of 26--28 mag and 21-24 mag, respectively,
generally rule out the existence of accretion disks, if considering X-ray 
emission from CCOs is powered by accreting from 
them \citep{wkc07,mig+08,del+11}. 

Deep \textit{Spitzer} imaging of
the CCOs RX~J0822.0$-$4300 (in the SNR Puppis A), 1E~1207.4$-$5209 (in the SNR
PKS~1209$-$52), and CXOU~J232327.8+584842 (in the SNR Cas A) were carried out
at 4.5~$\mu$m and 8.0~$\mu$m bands, with no MIR counterparts 
detected (\citealt{wkc07}; \citealt*{kcw13}). Comparing to the 4U~0142+61 dust disk case, 
\citet{wkc07} argued that if dust disks exist around them, because their
X-ray luminosities are two orders of magnitude lower than that of 4U~0142+61,
the non-detection could be due to their relatively weak irradiation fluxes.

\subsubsection{XDINSs}

There are seven XDINSs discovered by \textit{ROSAT} all-sky X-ray survey,
which have properties characteristic of a class of their own. They are
nearby (100--500\ pc) and young ($< 1$ Myr), and generally have only 
thermal emission.
Intensive studies of them were conducted, and they are thought, arguably,
to be linked to magnetars as their inferred magnetic fields are of the order
of 10$^{13}$~G and their luminosities are comparable or even greater than
their rotational energy loss rates (so-called spin-down luminosities).
For detailed reviews of these neutron stars, see \citet{hab07} 
and \citet{kap08}.

The ultraviolet/optical counterparts to the seven stars have all been
identified. However, the detections indicate that the emission is very
likely the Rayleigh-Jeans tail of the surface thermal emission, although
detailed properties are not fully understood (e.g., \citealt{kap+11}).
Deep searches for their IR counterparts were carried out, with surrounding disks
as the main targets \citep{loc+07,mig+08}. No counterparts were found
and the flux upper limits were 20--23 mag at NIR $H$ or $K_s$ bands. 
In addition, for the purpose
of searching for substellar companions to these neutron stars, \citet{pnh09} 
conducted similar searches and similar upper limits were obtained.
Given the low X-ray luminosities of XDINSs 
($L_{\rm X}\sim 10^{31}$--10$^{32}$ erg s$^{-1}$), \citet{loc+07} 
found that a hypothetical accretion disk would be bright due to 
internal viscous heating by using the fallback disk models given
by \citet{phn00}. The derived upper limits on the inflow rates of the
hypothetical disks were $\sim 10^{-10}\ M_{\odot}$~yr$^{-1}$.
It should be noted that due to their slow rotation and thus possibly large
inner disk radii (i.e., equal to either corotation radius 
or light cylinder radius; \citealt{loc+07}), the inflow-rate upper limits
are not very constraining. As a comparison, low-mass X-ray binaries 
known in our Galaxy, which include those that contain accreting neutron stars, 
have the order of magnitude mass-inflow rates in their accretion disks and
their disks are easily detectable at optical wavelengths.

In addition, \citet{pos+10} conducted a search for submm emission from
RX~J1856.5$-$3754, the brightest and closest member among the seven XDINSs.
No counterpart was found.  The flux upper limit was 5 mJy, setting an upper
limit of a few Earth masses on the cold, optically thin dust or a mass
accretion limit of $< 10^{14}$ g~s$^{-1}$ on the basis of the disk model
given by \citet{phn00}.

\subsection{Regular pulsars}
\label{subsec:rp}
Efforts have also been made to investigate the possible existence of debris
disks around regular radio pulsars, the major class among neutron
stars thus far known \citep{man+05}. 
Their main energy output is
a pulsar wind which is basically in a form of high-energy particles and
whose luminosity is limited by the spin-down luminosity $L_{\rm sd}$
of a pulsar.
As a result, the interaction would be between the pulsar wind and the
surrounding disk. Calculations about the possible interaction
processes were given \citep{ff96,jon07,jon08,cs08}.

With the purpose of searching for cold dust disks, several nearby 
(distances generally lower than 500 pc) regular pulsars were often 
included along 
with MSPs as the targets \citep{pc94,koc+02,lww04}. 
No counterparts
were found, and the obtained flux upper limits and mass limits on putative 
disks are similar to those given in Section~\ref{subsec:msp}. 
Additionally it is worth noting that one peculiar pulsar J1119$-$6127, 
which belongs to the 
so-called high magnetic-field pulsar group \citep{kas10}, was also
searched at near-IR bands but with no emission detected \citep{mpr+07}.

Recently \citet{dan+11} have found a candidate IR counterpart to the Vela pulsar
from \textit{Spitzer} imaging at 3.6 and 5.8~$\mu$m. If it is confirmed, 
although the magnetospheric origin for the the detected IR emission
could not be excluded based solely on the Vela's multiwavelength spectrum,
the existence of a debris disk possibility around this middle-aged 
(spin-down age is 11 kyr) pulsar was discussed in detail by \citet{dan+11}.
Such a case, if verified, would provide certain information for 
our understanding of the appearance of debris disks around regular radio
pulsars. The Vela pulsar 
has $L_{\rm sd}=6.9\times 10^{36}$ erg s$^{-1}$ and 
$L_{\rm X}\simeq 5.3\times 10^{32}$ \citep{pav+01} and is at a distance 
of 287~pc \citep{dod+03}, which imply fractions of 2.4$\times 10^{-8}$
or 3.2$\times 10^{-4}$ (at 5.8~$\mu$m) when its $L_{\rm sd}$ 
or X-ray emission
is considered, respectively. The value from considering X-ray heating 
is very similar
to that in the 4U~0142+61 case (see Figure~2), suggesting that X-ray 
heating probably plays the key role in making a debris disk detectable.
\begin{figure}[t]
\begin{center}
\includegraphics[scale=0.75]{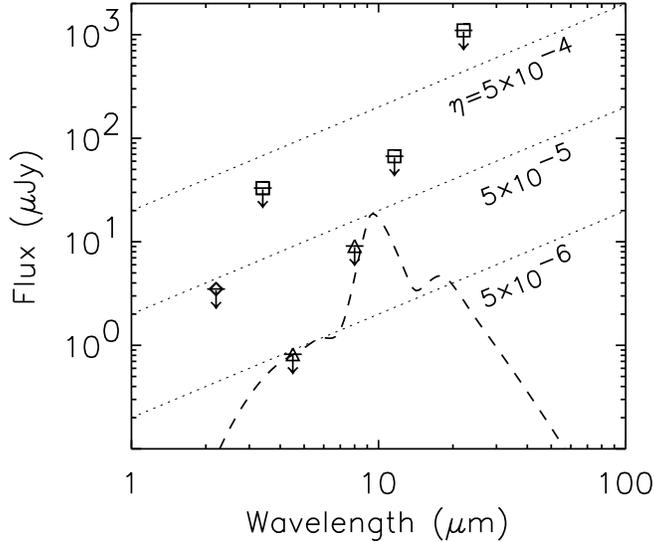}
\caption{Observed ground-based $K_s$ (diamond), \textit{Spitzer}\ (triangles), 
and WISE (squares) flux upper limits of PSR J0729$-$1448. Different fraction
values of $F_{\rm sd}$ are indicated by dotted lines. Emission from
a debris-disk model is displayed by the dashed curve.
}
\end{center}
\end{figure}

Having selected those relatively young (ages of several tens of kyr) and 
supposedly bright (i.e., high $L_{\rm sd}/d^2$ values, where $d$ is 
the estimated distance for a pulsar) as the targets among the known radio 
pulsars, \citet{wan+14} carried out ground-based
NIR $K$-band imaging of nearly 20 such pulsars. On the basis
of the results from $K$-band imaging, which helped further select out those 
pulsars in non-crowded fields as targets since source confusion 
at IR could be a severe
problem for counterpart identification, seven pulsars were imaged with 
\textit{Spitzer} at 4.5 and 8.0~$\mu$m. No IR counterparts were found.
The flux upper limits for PSR J0729$-$1448 are shown in Figure~3. This pulsar
has
$F_{\rm sd}=L_{\rm sd}/(4\pi d^2)=1.2\times 10^{-10}$ erg s$^{-1}$ cm$^{-2}$,
where $d=4.4$ kpc is used, and is taken as a typical example as other pulsar
targets have similar properties \citep{wan+14}.
The Wide-field Infrared Survey Explorer (WISE) band 1, 3 
and 4 (for WISE observations, see below) flux upper limits are also included.
Simply assuming a fraction $\eta$
of the energy flux from the pulsar would be re-radiated by a disk 
at the IR bands, $\eta F_{\rm sd}=\nu F_{\nu}$ where
$F_{\nu}$ is the flux density at a band of frequency $\nu$,
the upper limits reached $\eta\sim 5\times 10^{-4}$ at 20~$\mu$m and
$\eta \simeq 5\times 10^{-6}$ at 4.5 $\mu$m. The first fraction value is 
nearly as deep as that for the planetary system PSR B1257+12 obtained
from \textit{Spitzer} MIPS observations \citep{bry+06} and the latter
is at least one order of magnitude deeper than that of the disk case of 
the magnetar 4U~0142+61 ($\eta\sim 10^{-4}$; Figure~2). 
\citet{cs08} have proposed that X-ray emission from a neutron star
provides heating of a debris disk. If this is the case, assuming 
$L_{\rm X}\simeq 10^{-3} L_{\rm sd}$ (\citealt{bt97})
the fraction reached at 4.5~$\mu$m would be 5$\times 10^{-3}$, 
actually not sufficiently deep in comparison with the 4U~0142+61 (or
the Vela pulsar) case.

Launched in 2009 December, WISE 
mapped the entire sky at 3.4, 4.6, 12, and 22 $\mu$m (called W1, 
W2, W3, and W4 bands, respectively) in 2010 with FWHMs of 
6.1, 6.4, 6.5, and 12.0 arcsec  in the four bands, 
respectively (see \citealt{wri+10} for details). 
The WISE all-sky images and source catalogue were released in 2012 March.
Given its point-source detection limits of $\sim$0.08--6 mJy at the four 
bands \citep{wri+10}, WISE imaging could have provided a sensitive search 
for debris 
disks around all known neutron stars, although it suffers from source 
confusion and contamination, particularly at the Galactic plane, because of
its mediocre spatial resolution. Wang et al. (2014, in preparation) have
conducted searches for MIR counterparts to all radio pulsars using WISE images.
Preliminary results from the data analysis of the released WISE source
catalogue are negative, with no counterparts found. The 4.6~$\mu$m flux
upper limits, normalized by spin-down fluxes of pulsars, reached 
$\sim$10$^{-6}$.
Detailed image-data analyses for possible detections of MIR objects near 
pulsars' positions are underway, and approximately 30 such sources are found. 
In order to identify them, multiwavelength observations are needed.

\section{Discussion and Summary}
\label{dis}
Sufficient observational evidence has shown that debris disks are ubiquitous
around from different main-sequence stars (e.g., see \citealt{wya08} and 
references therein) to white dwarfs (e.g., \citealt{fjz09}; \citealt{xj12} 
and references therein). In main-sequence stars, debris disks may likely
be the remnants of protoplanetary disks at early ages of $\sim$10~Myr and
then are replenished by planetesimal collisions. In white dwarfs, which
like the neutron stars are also post main-sequence, compact stars, 
debris disks are
thought to be formed from material produced by tidal disruption
of planetary bodies \citep{gra+90,jur03} that have survived through late phases 
of stellar evolution \citep{ds02}. Comparing neutron stars
to them, the major uncertainties would be the initial conditions and heating
mechanisms if the current scenario of fallback disk formation is believed.
For young neutron stars (magnetars, CCOs, and XDINSs) and regular radio 
pulsars,
a debris disk would be the remnant of a fallback disk. The initial mass
of the fallback disk and its interaction with a newly-born neutron star 
are quite uncertain, which would affect subsequent evolution and 
detectability.
While dust grains around normal stars or white dwarfs are nicely heated to
100--1000~K temperature by ultraviolet/optical photons from host stars, 
revealing their existence, it is not clear how much a pulsar's particle wind 
can heat up a disk. 
Hopefully in the near future, with great capabilities of the next generation 
telescopes such as the thirty meter telescope and 
\textit{James-Webb Space Telescope}, these uncertainties would be cleared
out by detections or even sufficiently deep upper limits of a number of
different classes of neutron stars.

In summary, deep searches for disks around different types
of neutron stars have been carried out. For magnetars, 6 of them were found
with NIR emission, which was seen (except for 4U~0142+61) 
to be related to magnetars' outburst
activities. Further observations to identify the origin of the NIR emission is 
needed. Among the 6 magnetars with NIR counterparts, 4U~0142+61 has been
found to have a MIR component in its optical and IR broad-band spectrum
and it is likely
indicative of a dust disk around this X-ray emitting neutron star. Deep
MIR imaging of the magnetars 1E~2259+586 and 1E 1048.1$-$5937 detected the
first source at MIR 4.5~$\mu$m but did not detect the second one, and the
results suggest a similar dust disk to that of 4U~0142+61 around the first
source (although only on the basis of the MIR and NIR $K_s$ band detections) 
and exclude the existence of a similar disk around the second one. 
For CCOs and XDINSs, the deepest optical and NIR observations were carried 
out. The non-detections of them rule out accretion
(from a disk) as the power source to produce the observed X-ray emission from
the CCOs, while the upper limits on the XDINSs are not very constraining due
to possibly large inner radii considered for their putative accretion disks. 
Deep MIR imaging of CCOs was also conducted, but the obtained upper limits 
can not
conclusively determine whether or not dust disks exist. 
A few millisecond and middle-aged radio pulsars were searched.
The discovery of a candidate MIR counterpart
to the Vela pulsar, if confirmed, would provide a certain case for our
understanding of disk heating and detectability of disks around regular pulsars.
Extensive searches using WISE imaging data are underway, and hopefully
the results would bring to light the general existence of debris disks
and their evolution under pulsars' extreme environments.

We gratefully thank anonymous referees for constructive suggestions.
This research was supported by the National Natural Science Foundation of 
China (11073042, 11373055) and the Strategic Priority Research Program 
``The Emergence of Cosmological Structures" of the Chinese Academy 
of Sciences (Grant No. XDB09000000).  Z.W. is a Research Fellow of the 
One-Hundred-Talents project of Chinese Academy of Sciences.
\bibliographystyle{model2-names}

\end{document}